\documentclass[11pt,a4paper]{article}
\usepackage[margin=25mm]{geometry}
\usepackage{fontspec}
\usepackage{amsmath}
\usepackage{unicode-math}
\usepackage{graphicx,booktabs,tabularx,array}
\usepackage[section]{placeins}
\usepackage[font=small,labelfont=bf]{caption}
\usepackage[authoryear,round]{natbib}
\usepackage{xurl}
\usepackage[hidelinks,unicode]{hyperref}
\title{Empirical rather than simulated spectral band adjustment of MODIS to Sentinel-3 OLCI for European vegetation monitoring}
\author{Hongxiao Jin\\[0.4em]\small Department of Earth and Environmental Sciences, Lund University, Sweden}
\date{}
\begin{document}
\maketitle
\begin{abstract}
Spectral band adjustment derived from convolved spectra describes sensor differences under controlled conditions, but its applicability to operational MODIS and Sentinel-3 OLCI products requires assessment. We compare this approach with adjustment estimated directly from satellite observations for Medium Resolution Vegetation Phenology and Productivity (MR-VPP) 6.0. Both observation streams undergo atmospheric correction and Ross--Thick/Li--Sparse Reciprocal bidirectional reflectance distribution function (BRDF) normalization to nadir viewing and local solar noon before estimation. Using European observations from 2019--2024, we derive annual OLCI/MODIS reflectance-ratio relationships with unadjusted MODIS normalized difference vegetation index (NDVI). The operational correction uses quadratic red and linear near-infrared (NIR) functions. We compare these relationships with irradiance-weighted canopy simulations and assess the operational correction at 460 locations. Simulated and empirical relationships differ, particularly for red reflectance at high NDVI. Background-dependent simulated ratios also demonstrate the limitations of a single NDVI-based function. The empirical correction reduces red and NIR root-mean-square error (RMSE) at 66.3\% and 71.1\% of locations, with median paired reductions of 5.98\% and 13.67\%. Median paired RMSE decreases by 0.86\% for NDVI but increases by 0.95\% for the two-band enhanced vegetation index, 0.74\% for NIR reflectance of vegetation and 7.07\% for the Plant Phenology Index (PPI), calculated with stream-specific historical references. This study establishes an operational empirical adjustment and shows that the specified spectral simulation does not reproduce the relationship between the normalized products. It provides evidence for improved reflectance consistency while demonstrating that vegetation-index benefits require separate assessment.
\end{abstract}

\FloatBarrier

\section{Introduction}\label{sec:introduction}

Long satellite records of vegetation support studies of ecological seasonality \citep{TerasakiHart2025}, forest-cover change \citep{Hansen2013}, drought responses \citep{Jin2023} and longer-term environmental change \citep{Zhu2016}. Maintaining these records often requires combining successive satellite missions. However, corresponding bands on different instruments do not necessarily measure the same reflectance, even when observing the same surface \citep{Gonsamo2013,Trishchenko2002}. Differences introduced by a sensor transition can therefore be confounded with the vegetation changes that the record is intended to describe. Such discontinuities have been documented in long-term vegetation-index records and can affect trend interpretation \citep{Tian2015}.

Spectral band adjustment addresses differences associated with the spectral response functions (SRFs) of corresponding bands. A common approach applies the two sensor responses to the same hyperspectral reflectance spectrum, then fits a relationship between the resulting synthetic band values or their ratios. The input spectra may come from radiative-transfer simulations, measured spectral libraries or imaging spectroscopy. \citet{Trishchenko2002} quantified sensor-response effects using representative spectra and atmospheric simulations. \citet{Gonsamo2013} derived corrections for 21 sensors from simulated spectra and evaluated them with independent data, while \citet{Miura2013} used EO-1 Hyperion observations to examine compatibility across atmospheric-correction levels. \citet{VillaescusaNadal2019} compared adjustment formulations and found that the most effective model differed among bands. These studies establish a physical and methodological basis for adjustment, but do not imply a single formulation that can be transferred without assessment to another sensor pair and processing configuration.

A regression fitted to spectrally convolved data and a regression fitted directly to paired satellite observations estimate different relationships. In the former, both band values are generated from the same input spectrum; in the latter, they are retrieved independently by two observation systems. Regression is therefore a statistical tool shared by the approaches, rather than the feature that distinguishes them. Transformations derived directly from overlapping satellite observations, such as those reported for Landsat reflectance and normalized difference vegetation index (NDVI) by \citet{Roy2016}, provide an empirical route to product consistency. Their interpretation nevertheless requires attention to atmospheric correction, directional reflectance and sampling. \citet{Miura2013} showed that compatibility can depend on atmospheric processing, and \citet{Swinnen2022} found that the largest consistency improvement in their multi-mission assessment came from using a common atmospheric-correction method. The applicability of a spectrally derived relationship must therefore be examined at the processing level where it will be used.

For adjustment between the Moderate Resolution Imaging Spectroradiometer (MODIS) and OLCI, the scientific question is whether a relationship derived by applying the sensor responses to common spectra reproduces the relationship observed between the operational surface-reflectance products after major atmospheric and directional effects have been addressed. We examine this question using observations that first undergo atmospheric correction and then bidirectional reflectance distribution function (BRDF) normalization with the same Ross--Thick/Li--Sparse Reciprocal kernel model to nadir viewing and local solar noon. This common physical framework improves comparability before empirical fitting. The detailed retrieval implementations still differ, so the comparison evaluates the relationship between normalized products without assuming that all remaining differences arise from SRFs alone.

Here we develop and assess the MODIS-to-Sentinel-3 Ocean and Land Colour Instrument (OLCI) adjustment implemented in Medium Resolution Vegetation Phenology and Productivity (MR-VPP) 6.0. The application uses the Plant Phenology Index (PPI; \citealp{Jin2014}), which requires red and near-infrared (NIR) reflectances, motivating adjustment before index computation. Six overlapping years, 2019--2024, provide direct empirical relationships across European vegetation conditions. We compare their NDVI dependence with a controlled irradiance-weighted canopy simulation, quantify variation among annual empirical fits, and assess the operational equations at 460 European locations. Evaluating both reflectance and derived indices connects the physical comparison to the variables used in monitoring. The contribution is an empirical adjustment grounded in normalized satellite observations, an examination of whether the specified simulation describes that relationship, and evidence of the adjustment's application outcomes.

\FloatBarrier

\section{Satellite observations and comparability}\label{sec:satellite-observations-and-comparability}

\subsection{Reflectance products and reference geometry}\label{sec:reflectance-products-and-reference-geometry}

\begin{figure}[htbp]
\centering
\includegraphics[width=0.83\linewidth]{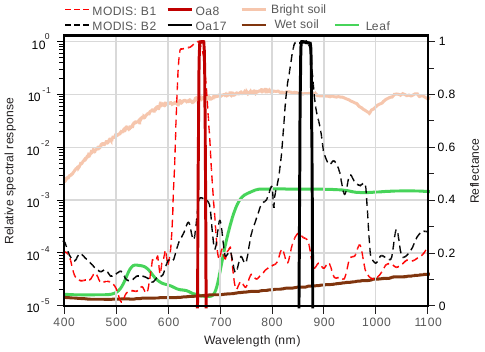}
\caption{Spectral response functions of the MODIS red and NIR bands (bands 1 and 2) and their Sentinel-3 OLCI counterparts (Oa08 and Oa17), shown with illustrative leaf, wet-soil and bright-soil reflectance spectra. Sensor responses use the logarithmic left axis, and surface reflectance uses the right axis. Differences in band position, width and response outside the main bandpass cause the sensors to weight the same spectrum differently. The leaf spectrum comes from the coupled PROSPECT/4SAIL modeling framework; the wet-soil spectrum is the PROSAIL reference, and the bright-soil spectrum is from the Johns Hopkins University (JHU)/Beckman Instruments spectral library (Entisol/Torripsamment, ID 0015). MODIS responses are available at \url{https://mcst.gsfc.nasa.gov/calibration/parameters}; OLCI-A responses are from \url{https://sentinels.copernicus.eu/documents/247904/2700530/S3A_OL_SRF_20160713_mean_rsr.nc4}. The JHU library description is available at \url{https://speclib.jpl.nasa.gov/documents/jhu_desc}.}\label{fig:1}
\end{figure}

The analysis uses MODIS and Sentinel-3 OLCI red and near-infrared (NIR) reflectance over Europe during their 2019--2024 overlap. MODIS is the source record to be adjusted, and OLCI defines the reference system. The selected bands are MODIS band 1 and OLCI Oa08 for red, and MODIS band 2 and OLCI Oa17 for NIR. Figure~\ref{fig:1} shows their spectral response functions alongside illustrative leaf and soil reflectance spectra. The bands sample corresponding spectral regions but weight the spectra differently, providing the physical motivation for the adjustment defined in Section 3.1.

MODIS reflectance was obtained from the MCD43A4 nadir BRDF-adjusted reflectance (NBAR) product Collection 6.1, with associated MCD43A2 quality information. The MODIS processing uses atmospherically corrected surface observations and a kernel-driven BRDF retrieval combining Terra and Aqua measurements \citep{Schaaf2002}. MCD43A4 is generated daily from a moving 16-day observation window, with the day of interest emphasized. Its nominal date therefore identifies a normalized retrieval rather than an instantaneous acquisition.

The OLCI inputs were project-supplied NBAR products generated using the ReBeLS approach \citep{LeonTavares2024}. For both sensor streams, atmospheric correction produces top-of-canopy (TOC) reflectance, followed by BRDF normalization using the Ross--Thick volumetric and Li--Sparse Reciprocal geometric kernels. In the implementation used here, both MODIS and OLCI reflectances were evaluated at nadir viewing and the solar zenith angle corresponding to local solar noon at each pixel and nominal date. This differs from the local 10:00 reference stated in \citet{LeonTavares2024}. The two datasets therefore share the same reference geometry for a given location and date, although their detailed BRDF inversion procedures differ. The empirical comparison consequently concerns atmospherically corrected surface reflectance normalized to a common viewing and illumination geometry.

The common BRDF model and reference geometry reduce differences associated with acquisition geometry before the band relationship is estimated. Atmospheric and BRDF retrieval uncertainties, temporal observation weights and spatial support can still differ between streams. We therefore interpret the fitted factors as an operational MODIS-to-OLCI mapping under the specified processing configuration, with the simulation providing a separate estimate of SRF effects under controlled conditions.

\subsection{Spatial alignment and calibration sampling}\label{sec:spatial-alignment-and-calibration-sampling}

\begin{figure}[htbp]
\centering
\includegraphics[width=0.94\linewidth]{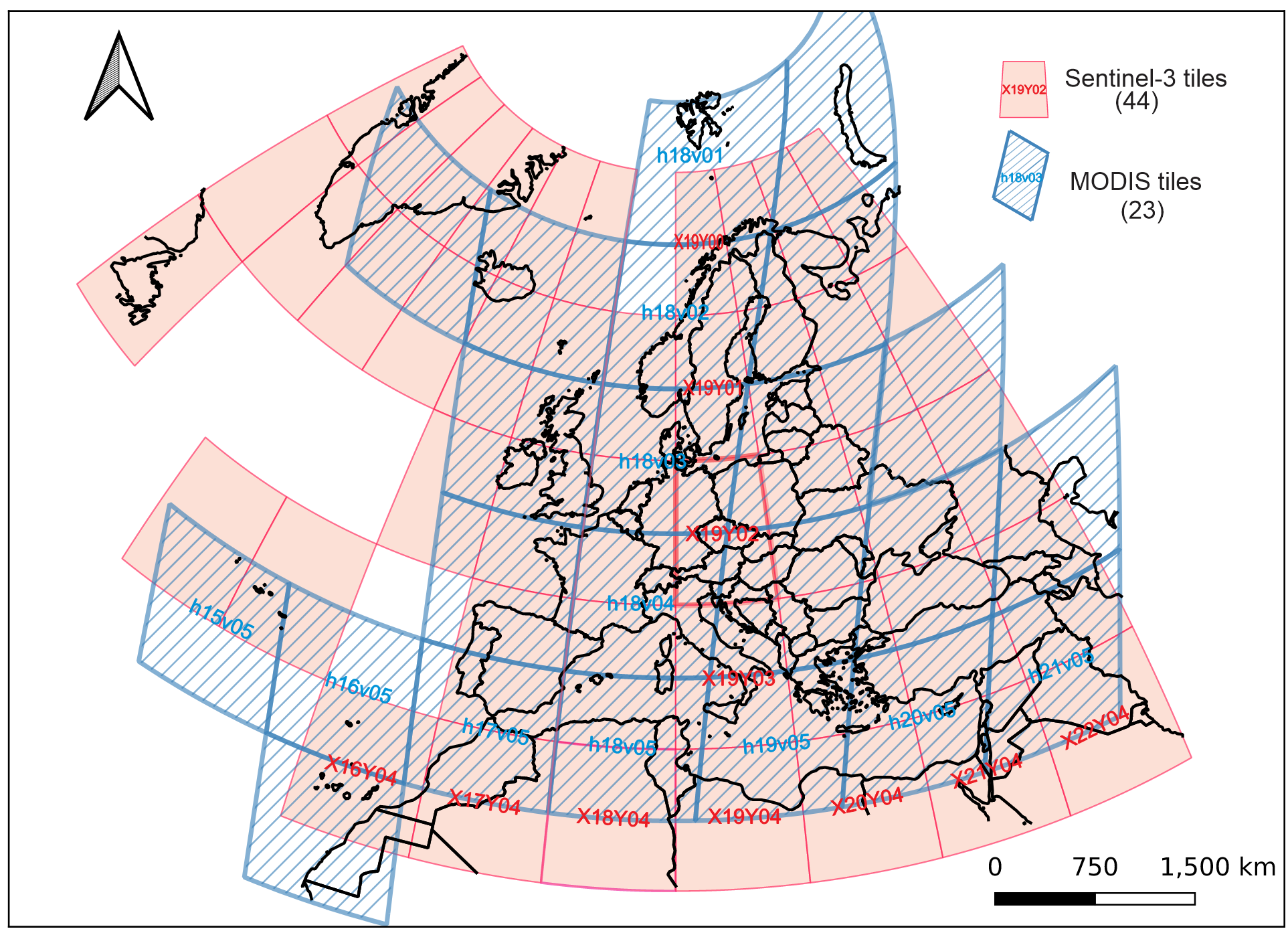}
\caption{Spatial coverage of the satellite tiles used to estimate the empirical MODIS-to-OLCI adjustment relationships over Europe during 2019--2024. Pink shading and red outlines indicate the 44 Sentinel-3 OLCI processing tiles; blue outlines and hatching indicate the 23 intersecting MODIS tiles. Selected tile identifiers are labelled for orientation.}\label{fig:2}
\end{figure}

The calibration domain comprises 44 OLCI processing tiles intersecting 23 MODIS tiles across Europe (Figure~\ref{fig:2}). MODIS reflectance and associated quality layers were mosaicked and reprojected from the native sinusoidal grid to the OLCI geographic grid (EPSG:4326) using nearest-neighbor resampling. The destination grid contains 3,360 × 3,360 cells per 10° tile, with an angular spacing of 1/336°. This alignment provides corresponding geographic samples while retaining the intrinsic spatial limitations of the source observations; it does not make the sensors' footprints identical.

Calibration samples were restricted to central cells of homogeneous 5 × 5 neighborhoods in the resampled C3S 2020 land-cover map, version 2.1.1. Built-up areas, bare areas, water, permanent snow and ice, and missing land-cover values were excluded. This selection reduces the influence of land-cover boundaries and spatial misregistration on the estimated relationship. Homogeneity refers to the mapped land-cover class and does not imply identical canopy or background spectra within the neighborhood.

Calibration observations were matched by the year and day of year encoded in the OLCI NBAR filenames. The nominal OLCI sequence, days 005, 010, 015, \ldots, 365, provides 73 target dates per year. MODIS products for those dates were assembled on the common grid. Paired samples were screened using the MODIS band-level BRDF quality information and OLCI quality flags, and fitting required finite reflectance strictly between zero and one in all four bands. Tile--date groups with fewer than 100 valid cells were omitted. Matching nominal product dates limits temporal separation but does not equalize the products' observation windows.

\subsection{Observations for application assessment}\label{sec:observations-for-application-assessment}

The implemented adjustment was assessed using a supplied list of 500 European locations. MODIS comparison-period inputs were unavailable for 40 locations, leaving 460 evaluable locations. These locations provide a geographically distributed application assessment; they are not documented as a probability sample or as excluded from the calibration data.

At each location, the assessment used a centered 3 × 3-cell patch on the project MODIS grid. OLCI values were selected using the cells containing the corresponding MODIS cell centers. This sampling differs from the homogeneous-neighborhood selection used for calibration. The patch provides a local spatial summary, and its nine cells are not treated as independent replicates.

The site comparison also uses a different temporal pairing procedure from calibration. Quality-screened OLCI red and NIR reflectances were linearly interpolated to MODIS production dates using valid endpoints no more than 10 days apart, without extrapolation. Reflectance adjustment and index calculation were performed per pixel before averaging over common valid cells. The detailed masks, index reference settings and agreement statistics are specified in the assessment methods.

\FloatBarrier

\section{Spectral adjustment methodology}\label{sec:spectral-adjustment-methodology}

The analysis uses two sources of calibration information within a common adjustment framework. The simulation applies both sensor SRFs to each modeled spectrum, whereas empirical estimation uses paired MODIS and OLCI NBAR products after the processing described in Section 2.1. Comparing their adjustment relationships tests whether the specified spectral representation describes the observed product relationship. The operational empirical equations are then assessed against unadjusted MODIS using OLCI as the reference.

\subsection{Definition and application of spectral band adjustment}\label{sec:definition-and-application-of-spectral-band-adjustment}

Spectral band adjustment expresses the reflectance measured by a source sensor in the spectral reference system of another sensor. Corresponding bands can yield different reflectances for the same surface because their spectral response functions (SRFs) weight different portions of its reflectance spectrum. The resulting difference depends on both the sensor responses and the observed spectral shape \citep{Gonsamo2013,Trishchenko2002}.

For the canopy simulations used here, synthetic band reflectance is calculated as

\begin{equation}
\rho_{s,b} = \frac{\int\rho(\lambda)\, E(\lambda)\, S_{s,b}(\lambda)\, d\lambda}{\int E(\lambda)\, S_{s,b}(\lambda)\, d\lambda},
\label{eq:1}
\end{equation}

where \(\rho(\lambda)\) is the surface reflectance spectrum, \(E(\lambda)\) is the prescribed incident spectral irradiance, and \(S_{s,b}(\lambda)\) is the SRF of band \(b\) on sensor \(s\). This normalized integration accounts for the spectral distribution of illumination as well as sensor sensitivity. The canopy and illumination assumptions are specified in Section 3.2.

We define the multiplicative spectral band adjustment factor as the ratio of reference-sensor to source-sensor band reflectance, following the formulation evaluated by \citet{VillaescusaNadal2019}. With MODIS as the source and OLCI as the reference,

\begin{equation}
f_{b} = \frac{\rho_{\mathrm{OLCI},b}}{\rho_{\mathrm{MODIS},b}}.
\label{eq:2}
\end{equation}

Here, \(b\) identifies the corresponding red or NIR bands. For a common simulated spectrum, Equation~\eqref{eq:2} describes their spectral-response difference under the prescribed illumination. When estimated from paired satellite products, the ratio instead describes the relationship between the processed reflectance streams and may include residual differences beyond their SRFs.

Routine application requires estimating the factor without a corresponding OLCI observation. We therefore parameterize it using NDVI calculated from the original MODIS red and NIR reflectances,

\begin{equation}
v = \frac{N - R}{N + R},
\label{eq:3}
\end{equation}

where \(R\) and \(N\) denote unadjusted MODIS red and NIR reflectance. NDVI-dependent spectral corrections provide a practical representation of variation along vegetation gradients \citep{Trishchenko2002,VillaescusaNadal2019}. NDVI does not uniquely determine the surface spectrum, however, so the adjustment remains an approximation across different canopy and background conditions \citep{Gonsamo2013}.

The adjusted reflectances are

\begin{equation}
\widehat{R} = R\,{\widehat{f}}_{R}(v),\quad\widehat{N} = N\,{\widehat{f}}_{N}(v),
\label{eq:4}
\end{equation}

where \({\widehat{f}}_{R}\) and \({\widehat{f}}_{N}\) are the estimated adjustment functions. A factor above one increases the source reflectance; a factor below one decreases it. The predictor is defined consistently during estimation and application using the original MODIS bands.

The correction is applied once to each valid source observation. Vegetation indices are subsequently calculated from the adjusted bands, following the band-first adjustment approach described by \citet{VillaescusaNadal2019}. This provides a common reflectance input for multiple indices while allowing the effect on each derived variable to be evaluated separately.

\subsection{Canopy simulation of MODIS–OLCI spectral differences}\label{sec:canopy-simulation-of-modisolci-spectral-differences}

A controlled canopy simulation was used to estimate the MODIS--OLCI spectral-response relationship independently of the two satellite retrieval chains. Both SRFs were applied to the same modeled surface spectrum, so each synthetic sensor pair shares the target, illumination and viewing conditions. This follows the spectral-convolution approach but permits a direct comparison with the relationship estimated from actual NBAR products. Leaf reflectance and transmittance were calculated with PROSPECT-D \citep{Feret2017} and supplied to 4SAIL to simulate canopy reflectance over a soil background, following the coupled leaf--canopy modeling framework reviewed by \citet{Jacquemoud2009}.

The full factorial design combined two contrasting soil spectra, 31 leaf area index (LAI) values, 13 solar zenith angles and nine prescribed diffuse-illumination fractions, producing 7,254 cases (Table~\ref{tab:1}). The wet and bright backgrounds were simulated separately and pooled for analysis. Their contrasting spectra extend the represented conditions from soil-dominated surfaces to dense canopies. Viewing was nadir, and leaf optical properties and canopy architecture were held fixed to isolate the effects of the varied parameters.

For each background, LAI and solar zenith angle, 4SAIL generated directional reflectance under direct and diffuse illumination. The canopy spectrum was constructed as

\begin{equation}
\rho(\lambda,\alpha) = (1 - \alpha)\rho_{\mathrm{dir}}(\lambda) + \alpha\rho_{\mathrm{dif}}(\lambda),
\label{eq:5}
\end{equation}

where \(\alpha\) is the prescribed diffuse fraction, applied uniformly across wavelengths. Solar zenith angle changes the canopy scattering geometry, while \(\alpha\) changes the relative contributions of direct and diffuse illumination.

Synthetic MODIS and OLCI red and NIR reflectances were calculated using the irradiance-weighted integration in Equation~\eqref{eq:1}. The reference irradiance was the Lawrence Berkeley National Laboratory AM1GH global-horizontal surface spectrum, generated with SMARTS 2.9.5 for clear-sky air-mass-one conditions (\url{https://coolcolors.lbl.gov/irradiance/index.html}). Its 5 nm samples were linearly interpolated to the common 1 nm grid used for the canopy spectra and sensor responses. Integration used the trapezoidal rule over the available 400--2500 nm spectral domain. A fixed irradiance spectral shape was retained across canopy solar angles and diffuse fractions; these parameters were varied independently as controlled sensitivity factors rather than coupled through an atmospheric simulation.

For each case, MODIS NDVI and OLCI-to-MODIS band ratios were calculated using Equations (2)--(3). To examine whether NDVI alone provides a compact description of the simulated ratios, a quadratic was fitted separately for each band:

\begin{equation}
{\widehat{f}}_{b,\mathrm{sim}}(v) = c_{0,b} + c_{1,b}v + c_{2,b}v^{2}.
\label{eq:6}
\end{equation}

Ordinary least squares assigned equal weight to all 7,254 cases, giving each background equal total weight. The factorial design does not produce equal sampling density across NDVI intervals. The fitted curves were assessed against the simulated points using ratio residuals and the reflectance differences obtained after applying the factors. These diagnostics quantify the adequacy of the quadratic approximation within the specified simulation ensemble. Simulated points and fitted curves are presented together in the Results.

\begin{table}[htbp]
\caption{\textbf{Parameter settings for the PROSPECT-D/4SAIL simulation.}}\label{tab:1}
\centering
\small
\renewcommand{\arraystretch}{1.15}
\begin{tabularx}{\linewidth}{@{}>{\raggedright\arraybackslash}p{0.32\linewidth}>{\raggedright\arraybackslash}X@{}}
\toprule
\textbf{Parameter} & \textbf{Setting} \\
\midrule
Soil background & Two fixed reference spectra: wet and bright \\
Leaf area index (LAI) & 0--6, increment 0.2 (31 values) \\
Solar zenith angle & 0--60°, increment 5° (13 values) \\
Prescribed diffuse fraction, \(\alpha\) & 0.1--0.9, increment 0.1 (nine values) \\
Total combinations & 2 × 31 × 13 × 9 = 7,254 \\
Leaf structure parameter, \(N\) & 1.5 \\
Chlorophyll a + b, \(C_{ab}\) & 40 \(\mu\mathrm{g}\,\mathrm{cm}^{-2}\) \\
Carotenoids, \(C_{ar}\) & 8 \(\mu\mathrm{g}\,\mathrm{cm}^{-2}\) \\
Brown pigments / anthocyanins & 0 / 0 \\
Equivalent water thickness, \(C_{w}\) & 0.015 cm \\
Dry matter per leaf area, \(C_{m}\) & 0.009 \(\mathrm{g}\,\mathrm{cm}^{-2}\) \\
Leaf-angle distribution & Campbell; mean inclination 57°; 18 angular bins \\
Hotspot parameter & 0.01 \\
View zenith angle & 0° \\
Solar / view azimuth & 0° / 0° \\
Spectral integration & 400--2500 nm; common 1 nm sampling; trapezoidal rule \\
Reference surface irradiance & LBL AM1GH global-horizontal spectrum; fixed spectral shape \\
\bottomrule
\end{tabularx}
\end{table}

The spectral grid denotes numerical sampling, not the original measurement resolution of the input spectra. MODIS response contributions below 400 nm were excluded because corresponding target spectra were unavailable. They account for approximately 0.011\% and 0.028\% of the solar-weighted response area for red and NIR, respectively; these fractions do not directly quantify reflectance error.

\subsection{Empirical estimation and operational adjustment equations}\label{sec:empirical-estimation-and-operational-adjustment-equations}

Empirical adjustment functions were estimated directly from the quality-screened MODIS--OLCI NBAR pairs described in Section 2.2. Atmospheric correction and the common Ross--Thick/Li--Sparse Reciprocal normalization precede this estimation. Unlike the simulation, each calibration pair therefore comprises two independently retrieved satellite reflectances rather than two integrations of one input spectrum. MODIS NDVI was calculated from the unadjusted red and NIR bands using Equation~\eqref{eq:3}, and the OLCI-to-MODIS reflectance ratio was calculated separately for each band using Equation~\eqref{eq:2}.

For each year from 2019 to 2024, ordinary least squares fitted a quadratic relationship between the band ratio and source MODIS NDVI, following the NDVI-dependent SBAF formulation evaluated by \citet{VillaescusaNadal2019}:

\begin{equation}
\frac{\rho_{\mathrm{OLCI},b,i,y}}{\rho_{\mathrm{MODIS},b,i,y}} = a_{0,b,y} + a_{1,b,y}v_{i,y} + a_{2,b,y}v_{i,y}^{2} + \varepsilon_{b,i,y},
\label{eq:7}
\end{equation}

where \(i\) indexes valid calibration samples, \(y\) denotes year, and \(b\) identifies red or NIR. All retained cells and dates across the 44 tiles were pooled within each year. Each sample received equal regression weight, without balancing the contributions of tiles or land-cover classes. The fitting objective was squared error in reflectance ratios; agreement in reflectance units was assessed separately.

To obtain fixed coefficients for application throughout the MODIS record, each annual coefficient vector was assigned equal weight:

\begin{equation}
{\bar{a}}_{k,b} = \frac{1}{6}\sum_{y = 2019}^{2024}a_{k,b,y},\quad\quad k \in \left\{ 0,1,2 \right\}.
\label{eq:8}
\end{equation}

This averaging gives each year the same contribution, irrespective of its number of retained observations. Annual variation was summarized by evaluating the six fitted functions at common NDVI values and reporting their minimum--maximum envelope. This envelope describes interannual variation in the fitted relationships, rather than a statistical confidence interval.

The operational red factor retains all three mean coefficients, whereas the NIR factor retains the mean intercept and linear coefficient:

\begin{equation}
{\widehat{f}}_{R}(v) = 1.063247 + 0.255060v - 0.471804v^{2},
\label{eq:9}
\end{equation}

\begin{equation}
{\widehat{f}}_{N}(v) = 1.044383 + 0.007799v.
\label{eq:10}
\end{equation}

The NIR expression sets the mean quadratic coefficient to zero while retaining the other coefficients from the annual quadratic fits; it was not obtained by a separate linear refit. Accordingly, the annual NIR envelope describes the full quadratic fits, not uncertainty around the operational linear function. The performance of the implemented equations is assessed using the paired reflectance and vegetation-index comparisons described in the following section.

Application follows Equation~\eqref{eq:4}: calculate NDVI once from the original MODIS bands, evaluate both factors using that NDVI, and multiply each original band by its corresponding factor. OLCI observations are required for calibration but not for subsequent application. The same fixed functions are used across years, and vegetation indices are calculated from the adjusted reflectances. These coefficients apply to the input processing configuration specified in Section 2.

\subsection{Assessment of reflectance and vegetation-index agreement}\label{sec:assessment-of-reflectance-and-vegetation-index-agreement}

The assessment examined whether the empirically derived operational adjustment improves MODIS red and NIR agreement with OLCI, and whether those benefits extend to derived vegetation indices. Unadjusted MODIS and the operationally adjusted MODIS products were compared with the same OLCI-derived reference at the locations and dates described in Section 2.3 during 2019--2024. This evaluates agreement between normalized products under the tested processing configuration. The simulation is compared with the empirical relationships in factor space; the satellite assessment does not test a simulation-derived correction against the empirical correction or establish accuracy against an independent surface reference.

\textbf{Paired observations.} Reflectances were converted to fractional units, and fill values were removed before analysis. Comparison screening retained MODIS red and NIR band-level QA codes 0 or 1 and OLCI quality flag quil = 4. The temporal interpolation described in Section 2.3 used the same valid OLCI endpoint dates for both bands, preserved exact date matches and did not extrapolate beyond available observations.

For each variable, the unadjusted and adjusted MODIS values and their OLCI reference were restricted to common valid cells and dates. Indices were calculated per pixel before averaging over the common cells in each 3 × 3 patch. A date was retained when at least one common valid cell was available. Site statistics were calculated across the resulting paired patch means. PPI-specific invalidity did not remove observations from the other variables, and observations were not excluded because their intersensor residuals were large.

\textbf{Vegetation indices.} NDVI was calculated using Equation~\eqref{eq:3}, with the bands of the corresponding observation stream. The NDVI recalculated from adjusted bands is distinct from the original MODIS NDVI used to predict the adjustment factors. EVI2 and NIRv were calculated as

\begin{equation}
\mathrm{EVI2} = 2.5\frac{N - R}{N + 2.4R + 1},
\label{eq:11}
\end{equation}

\begin{equation}
\mathrm{NIRv} = N\frac{N - R}{N + R},
\label{eq:12}
\end{equation}

following \citet{Jiang2008} and \citet{Badgley2017}, respectively. Each index was derived separately from unadjusted MODIS, adjusted MODIS and the temporally matched OLCI bands.

PPI was calculated using the project implementation of \citet{Jin2014}, with the historical red--NIR difference reference estimated separately for each observation stream. For each pixel, the 95th percentile of the native-date NIR-minus-red difference was bounded to 0.18--0.99 and increased by 0.005 to obtain the maximum-difference reference. MODIS references used the available 2000--2026 history; OLCI references used native 2019--2024 observations before interpolation. Historical MODIS reference estimation retained QA codes 0--3, whereas the paired comparison retained codes 0--1. PPI comparisons therefore assess the index with its stream-specific historical parameterization as well as its current reflectance inputs.

\textbf{Agreement statistics.} For each site and variable, signed bias and root mean square error (RMSE) were calculated as

\begin{equation}
\mathrm{Bias} = \frac{1}{n}\sum_{t = 1}^{n}\left( X_{M,t} - X_{O,t} \right),
\label{eq:13}
\end{equation}

\begin{equation}
\mathrm{RMSE} = \sqrt{\frac{1}{n}\sum_{t = 1}^{n}\left( X_{M,t} - X_{O,t} \right)^{2}},
\label{eq:14}
\end{equation}

where \(X_{M,t}\) is the unadjusted or adjusted MODIS patch mean, \(X_{O,t}\) is the corresponding OLCI-derived patch mean, and \(n\) is the number of common dates. Positive bias indicates higher MODIS values. Bias improvement was defined as a reduction in absolute bias. The paired percentage change in RMSE was

\begin{equation}
\Delta \mathrm{RMSE}(\%) = 100\frac{{\mathrm{RMSE}}_{\mathrm{adjusted}} - {\mathrm{RMSE}}_{\mathrm{unadjusted}}}{{\mathrm{RMSE}}_{\mathrm{unadjusted}}}.
\label{eq:15}
\end{equation}

Negative values indicate improvement. Percentage changes require a positive baseline RMSE. Across locations, each evaluable site received equal weight. Summaries report the proportion of sites improving and the median of the site-level paired percentage changes, rather than the percentage change between population medians. These descriptive summaries characterize variation in application outcomes without treating pixels or successive dates as independent replicates. Because exclusion of assessment locations from calibration is not documented, the comparison is not presented as independent validation.

\FloatBarrier

\section{Results}\label{sec:results}

\subsection{Simulated and empirical adjustment relationships}\label{sec:simulated-and-empirical-adjustment-relationships}

\begin{figure}[htbp]
\centering
\includegraphics[width=1.0\linewidth]{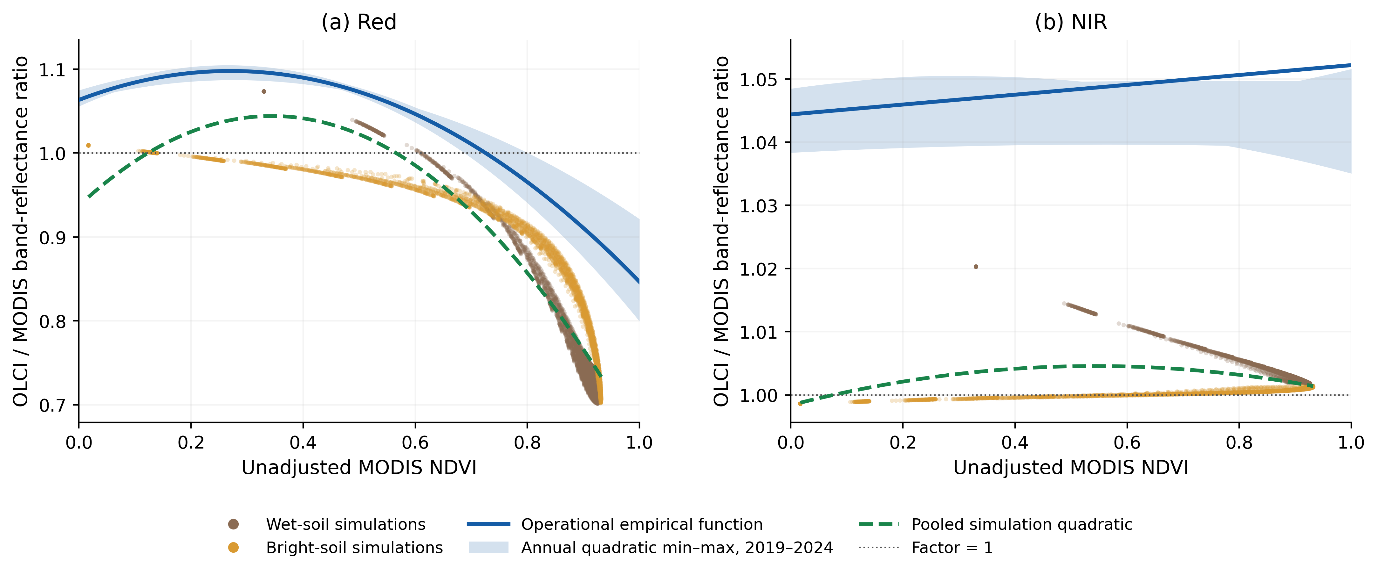}
\caption{\textbf{Simulated and empirical MODIS-to-OLCI adjustment relationships.} Red and NIR panels show OLCI/MODIS band-reflectance ratios against unadjusted MODIS NDVI. Points represent the 7,254 irradiance-weighted PROSPECT-D/4SAIL cases, with the two soil backgrounds distinguished by color. Solid blue curves show the operational empirical functions; blue shading shows the minimum--maximum envelope of the six annual empirical quadratic fits for 2019--2024. Dashed green curves show quadratics fitted to the pooled simulation cases and are displayed only over their NDVI range. The horizontal reference line denotes a factor of one. The annual envelope represents interannual fitted variation, not a confidence interval; for NIR it refers to full quadratic fits rather than the operational linear function.}\label{fig:3}
\end{figure}

The relationship estimated directly from the normalized satellite products differed from the relationship obtained by convolving the sensor responses with modeled spectra (Figure~\ref{fig:3}). Both showed stronger NDVI dependence in red than in NIR, but the adjustment magnitudes and background dependencies differed. The irradiance-weighted simulation covered source MODIS NDVI from 0.017 to 0.932. Its red ratios decreased markedly toward the upper end of this range, whereas NIR ratios remained closer to unity. The two soil backgrounds produced distinct relationships, particularly in NIR, demonstrating that a given NDVI did not uniquely determine the adjustment factor within the simulated ensemble.

The empirical red function increased at low NDVI, reached a maximum near NDVI 0.27 and then decreased. At NDVI 0.2 and 0.8, the operational red factors were 1.095 and 0.965, respectively. Thus, the implemented function increases red reflectance at low-to-moderate NDVI and decreases it at high NDVI. The operational NIR factor was weakly dependent on NDVI and remained above unity, increasing from 1.046 at NDVI 0.2 to 1.051 at NDVI 0.8. These factors correspond to NIR increases of approximately 4.6--5.1\% over that interval.

Annual empirical fits also differed in their predicted factors. At NDVI 0.8, the six red factors ranged from 0.940 to 1.000, while the full annual NIR quadratic fits ranged from 1.039 to 1.050. These ranges describe variation among the annual fitted relationships at a common vegetation-index value. The operational red function averages the annual quadratics; the operational NIR function additionally omits the mean quadratic contribution, as specified in Section 3.3. The annual NIR envelope therefore does not define uncertainty bounds for the operational linear factor.

The empirical and simulated results differed most clearly in the magnitude of the high-NDVI red adjustment and the positive NIR adjustment inferred from satellite observations. For example, at NDVI 0.8, the pooled simulation quadratics predicted red and NIR factors of 0.857 and 1.003, compared with operational empirical factors of 0.965 and 1.051. The simulated values in this comparison are predictions of the fitted quadratics, not individual simulated ratios or background-specific averages. Figure~\ref{fig:3} retains the simulated points to show the differences between those relationships and their pooled approximation.

The pooled quadratic did not reproduce the simulated ratios consistently across backgrounds. When applied to the same simulation cases, it increased red reflectance RMSE from 0.004985 to 0.007838 and NIR RMSE from 0.000776 to 0.001253. This result identifies a limitation of the NDVI-based statistical approximation within the controlled ensemble. It is distinct from the discrepancy between the empirical factors and the simulated points themselves. Figure~\ref{fig:3} therefore shows both the point relationships and the fitted simulation summaries; neither the synthetic fit diagnostics nor the curve comparison constitutes a satellite performance test of the simulation-derived adjustment. Sections 4.2--4.3 report the operational empirical adjustment's application outcomes.

\subsection{Reflectance agreement across European locations}\label{sec:reflectance-agreement-across-european-locations}

\begin{figure}[htbp]
\centering
\includegraphics[width=1.0\linewidth]{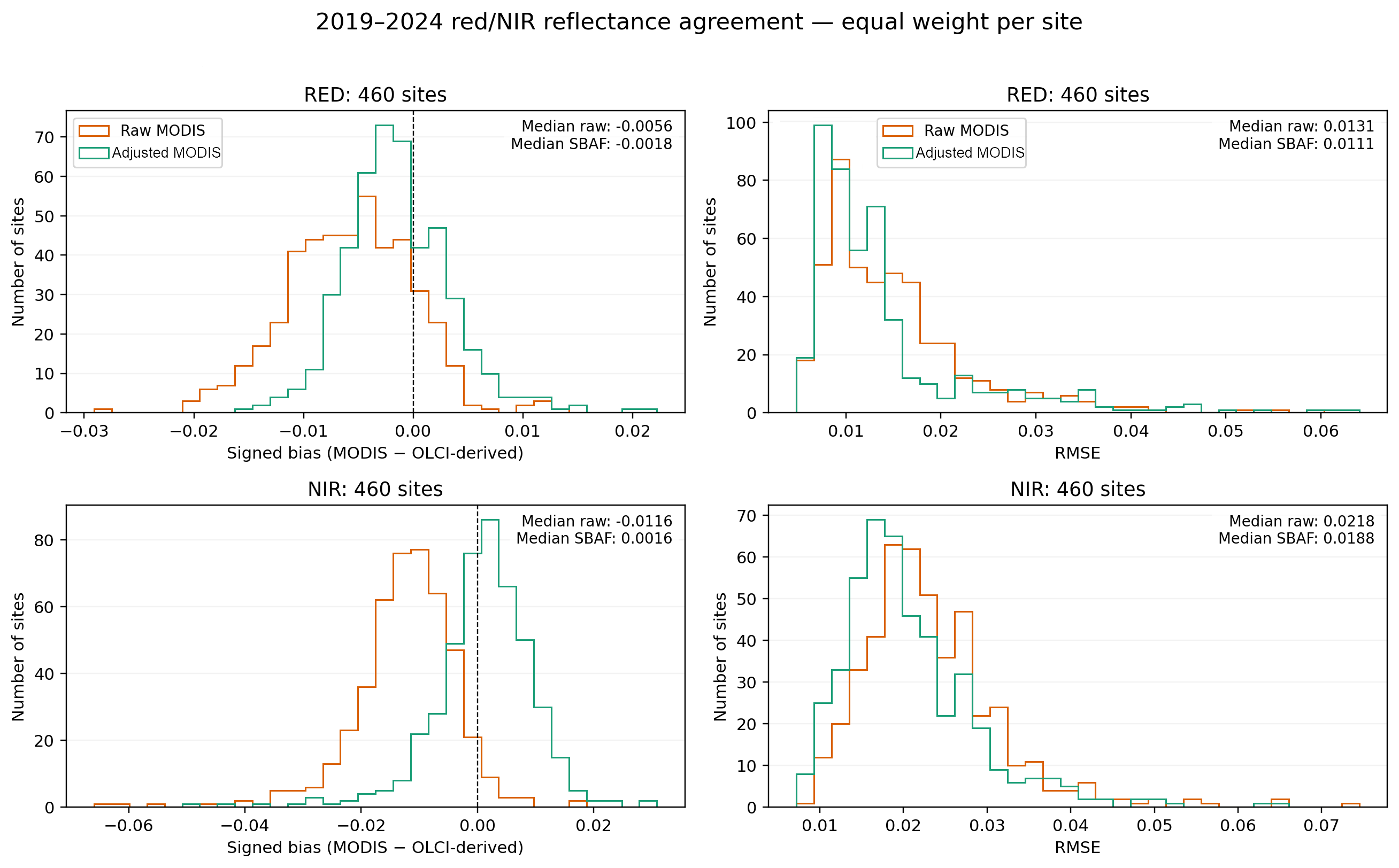}
\caption{\textbf{Reflectance agreement before and after operational SBAF adjustment.} Distributions of signed site bias (left) and RMSE (right) for red (top) and NIR (bottom), calculated over 2019--2024 from the paired 3 × 3-patch comparisons. Orange denotes unadjusted MODIS and green the operationally adjusted MODIS product. Each of the 460 evaluable sites contributes one statistic per band and processing stream. Dashed vertical lines in the bias panels indicate zero bias. The before-and-after distributions are descriptive; paired improvement rates are reported in Table~\ref{tab:2}. Values are in fractional reflectance units.}\label{fig:4}
\end{figure}

The operational empirical adjustment improved MODIS--OLCI reflectance agreement at most of the 460 evaluable locations during 2019--2024 (Figure~\ref{fig:4}; Table~\ref{tab:2}). Absolute site bias decreased at 345 locations (75.0\%) for red and 356 locations (77.4\%) for NIR. The improvement therefore extended beyond a small set of illustrative sites, although it was not universal.

Before adjustment, median site bias was negative in both bands, indicating lower MODIS reflectance relative to the OLCI-derived reference. The median red bias shifted from −0.00565 to −0.00183, while the median NIR bias shifted from −0.01156 to +0.00156. The adjustment substantially reduced the typical negative offset, with the NIR median crossing slightly above zero. The bias distributions nevertheless retained both negative and positive residuals after adjustment.

RMSE decreased at 305 locations (66.3\%) for red and 327 locations (71.1\%) for NIR. Median site RMSE changed from 0.01313 to 0.01107 in red and from 0.02181 to 0.01883 in NIR. Calculated from each site's paired before-and-after metrics, the median RMSE reductions were 5.98\% and 13.67\%, respectively. Thus, NIR showed a larger typical relative reduction, while retaining a higher median absolute RMSE than red.

For both bands, reductions in absolute bias occurred more frequently than reductions in RMSE. The correction therefore improved the mean offset at more locations than it improved overall agreement across dates. Remaining temporal disagreement and the locations without improvement show that the fixed adjustment does not reconcile the two reflectance streams equally well under all sampled conditions. Whether the band-level benefits are retained in vegetation indices is examined in Section 4.3.

\begin{table}[htbp]
\caption{\textbf{Site-level reflectance agreement before and after operational SBAF adjustment, 2019--2024.} Bias is MODIS minus the OLCI-derived reference. All reflectance values are in fractional units; each summary uses 460 evaluable sites.}\label{tab:2}
\centering
\small
\renewcommand{\arraystretch}{1.15}
\begin{tabularx}{\linewidth}{@{}>{\raggedright\arraybackslash}Xrr@{}}
\toprule
\textbf{Statistic} & \textbf{Red} & \textbf{NIR} \\
\midrule
Median signed bias, unadjusted & −0.00565 & −0.01156 \\
Median signed bias, adjusted & −0.00183 & +0.00156 \\
Sites with lower absolute bias & 345/460 (75.0\%) & 356/460 (77.4\%) \\
Median RMSE, unadjusted & 0.01313 & 0.02181 \\
Median RMSE, adjusted & 0.01107 & 0.01883 \\
Sites with lower RMSE & 305/460 (66.3\%) & 327/460 (71.1\%) \\
Median paired RMSE change & −5.98\% & −13.67\% \\
\bottomrule
\end{tabularx}
\end{table}

Median bias and RMSE summarize the site distributions. Median paired RMSE change is the median of the individual site percentage changes defined in Equation~\eqref{eq:15}, not the percentage change between the two reported median RMSE values.

\subsection{Vegetation-index responses}\label{sec:vegetation-index-responses}

\begin{figure}[htbp]
\centering
\includegraphics[width=1.0\linewidth]{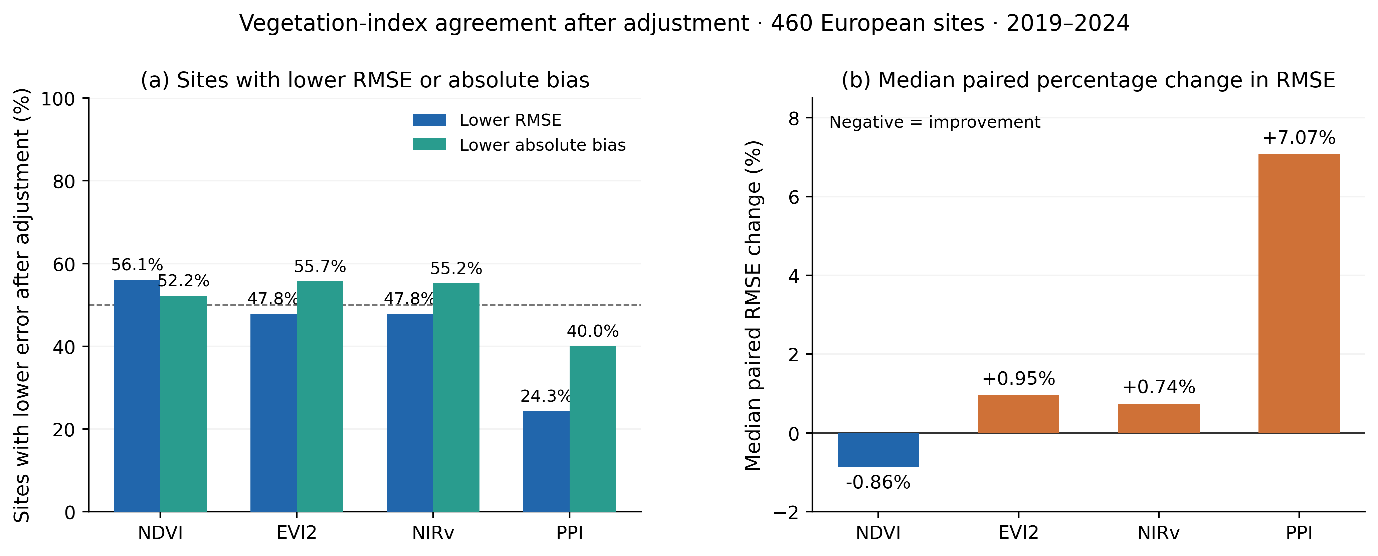}
\caption{\textbf{Effects of the operational SBAF adjustment on vegetation-index agreement across 460 European sites during 2019--2024.} (a) Percentage of sites with lower RMSE or absolute bias after adjustment; the dashed line marks 50\%. (b) Median paired percentage change in site-level RMSE, calculated as 100 × (adjusted − unadjusted)/unadjusted; negative values indicate improvement. Unadjusted and adjusted MODIS indices were compared with the same OLCI-derived reference using common valid observations within each index. PPI was calculated with dataset-specific historical reference parameters.}\label{fig:5}
\end{figure}

The reflectance improvements reported in Section 4.2 did not translate into similarly widespread improvements in vegetation indices (Figure~\ref{fig:5}). Across the 460 evaluable locations, NDVI RMSE decreased at 56.1\% of sites, but its median paired reduction was only 0.86\%. EVI2 and NIRv RMSE each decreased at 47.8\% of sites, with median paired increases of 0.95\% and 0.74\%, respectively. Their typical RMSE changes were therefore small, with benefits varying among locations.

Absolute bias and RMSE gave different assessments of the index responses. Absolute bias decreased at 52.2\% of sites for NDVI, 55.7\% for EVI2 and 55.2\% for NIRv. For EVI2 and NIRv, the adjustment thus reduced the mean offset at a modest majority of sites without reducing RMSE at a majority. These differences show that the effect of harmonization depends on the aspect of agreement being assessed.

PPI showed the least frequent improvement under the reference configuration specified in Section 3.4. Its RMSE decreased at 24.3\% of sites and absolute bias at 40.0\%, while the median paired RMSE increased by 7.07\%. Because PPI used separately estimated historical references for each observation stream, this outcome describes the combined reflectance-and-reference calculation. The comparison does not isolate the contribution of the current reflectance adjustment from that of the historical reference parameterization.

Taken together, the band and index comparisons establish a variable-specific outcome: the implemented adjustment improves red and NIR agreement at most assessed locations, while improvements in derived indices are less consistent. The observed index responses provide no basis for assuming an equivalent benefit in phenological dates, seasonal productivity or drought indicators, which were not evaluated by these agreement statistics.

\FloatBarrier

\section{Discussion}\label{sec:discussion}

\subsection{From spectral convolution to empirical product adjustment}\label{sec:from-spectral-convolution-to-empirical-product-adjustment}

The central comparison concerns the relationship between adjustment derived from common spectra and adjustment estimated directly from normalized satellite observations. Spectral convolution isolates the effect of sensor responses for the spectra and illumination supplied to the calculation. Empirical estimation instead describes the independently retrieved MODIS and OLCI products that must be reconciled in the operational record. The discrepancy in Figure~\ref{fig:3} shows that the specified simulation does not reproduce the observed NBAR relationship. Establishing that distinction is useful because a plausible spectral calculation and a well-defined regression do not, by themselves, establish an adequate adjustment for the products being used.

The common processing framework is important to this interpretation. Both observation streams first undergo atmospheric correction and then use the Ross--Thick/Li--Sparse Reciprocal BRDF model to estimate nadir reflectance at local solar noon. The empirical relationship is consequently estimated after the main atmospheric and acquisition-geometry effects have been addressed, rather than allowing those effects to enter an untreated band comparison. This makes the comparison with surface-spectral simulations more informative. Differences in the detailed retrieval algorithms, temporal weighting and spatial support remain, however, and cannot be partitioned from the fitted ratios alone.

NDVI dependence and spectral information. Canopy development changes red absorption, NIR scattering and background visibility, making NDVI dependence physically plausible \citep{Trishchenko2002,VillaescusaNadal2019}. NDVI nevertheless compresses two band measurements into one normalized ratio. Similar NDVI can arise from spectra with different magnitudes, red-edge shapes or background contributions, which need not yield the same intersensor band ratio. \citet{Gonsamo2013} used additional spectral information in their reflectance models. The separate background-dependent relationships here demonstrate why NDVI is a practical predictor but does not uniquely specify the spectral adjustment.

Physical simulation and statistical approximation. Two issues must be distinguished when interpreting Figure~\ref{fig:3}. First, the represented spectra and illumination determine the simulated point relationships. Second, the fitted quadratic compresses those relationships into one function. Pooling backgrounds can produce approximation errors even when the spectra and their integrations are internally consistent. The increased synthetic reflectance RMSE after adjustment demonstrates this limitation for the present ensemble. A more flexible predictor could reduce that approximation error, but its suitability for satellite products would still depend on the represented conditions. \citet{VillaescusaNadal2019} likewise found that effective adjustment formulations differed among bands.

The fitting objective introduces a further distinction. For positive source reflectance \(x_{i}\), reference reflectance \(y_{i}\) and predicted factor \(f\left( v_{i} \right)\), squared ratio error satisfies

\begin{equation}
\sum_{i}^{}\left\lbrack \frac{y_{i}}{x_{i}} - f\left( v_{i} \right) \right\rbrack^{2} = \sum_{i}^{}\frac{\left\lbrack y_{i} - x_{i}f\left( v_{i} \right) \right\rbrack^{2}}{x_{i}^{2}}.
\label{eq:16}
\end{equation}

Thus, fitting ratios by ordinary least squares is equivalent to weighting squared reflectance residuals by inverse squared source reflectance. Darker source observations receive greater weight for a given absolute residual, which is particularly relevant in red. This explains why the fitting objective differs from the unweighted reflectance RMSE used in the application assessment. It does not identify the cause or magnitude of the observed simulation--empirical discrepancy. Both the representation of the calibration data and the objective used to summarize them matter when selecting an operational function.

What the simulation represents. The two-background design spans near-zero to high NDVI while controlling the varied parameters. Leaf biochemistry and canopy architecture remain fixed, and two backgrounds do not represent the full diversity of European surfaces. Irradiance weighting accounts for the prescribed illumination spectrum in band integration, but its fixed spectral shape and independently varied solar angle and diffuse fraction do not reproduce their atmospheric coupling. The experiment therefore tests a specified physical representation. Its disagreement with the empirical relationship identifies a need to assess transfer to observed NBAR products; it does not establish that spectral-convolution methods generally lack practical value.

Implications for operational adjustment. The empirical functions use the actual MODIS and OLCI streams, their overlap and their common normalization framework, making them directly applicable to the configuration for which they are estimated. Their practical support comes from improved reflectance agreement at most assessed locations, not from assuming that empirical estimation is intrinsically superior to physical simulation. Conversely, fitting a regression to simulated or measured hyperspectral spectra does not make it equivalent to fitting paired satellite NBAR observations. The provenance of the calibration pairs must be stated when an equation is presented as an SBAF.

The present evidence supports using the empirical mapping for this processing configuration and evaluating its consequences for the intended variables. A comparison of both candidate corrections on the same withheld satellite observations would provide stronger evidence about their relative application performance. Such a test is not supplied by the current curve comparison and before-and-after assessment. Variation among annual empirical fits also reflects retained samples and observation conditions as well as possible changes in the product relationship, so it should not be attributed solely to sensor drift. The index responses in Section 4.3 show why the empirical mapping's band-level benefit must be followed through the application calculation.

\subsection{Why reflectance improvements need not improve vegetation indices}\label{sec:why-reflectance-improvements-need-not-improve-vegetation-indices}

The different outcomes for reflectance and vegetation indices can be understood through the joint red--NIR relationship and the parameters used to calculate each index. Band-level harmonization and index-level consistency are related objectives, but improvement in one does not establish improvement in the other. Sections 4.2--4.3 show widespread reductions in red and NIR disagreement, small typical EVI2 and NIRv RMSE changes, and more frequent PPI deterioration. These results define the extent of the demonstrated application benefit; the mechanisms below explain possible pathways without assigning their relative contributions from the aggregate statistics alone.

\textbf{Joint band errors.} Let \(e_{R}\) and \(e_{N}\) denote MODIS-minus-OLCI reflectance differences. For a differentiable index \(I(R,N)\), a first-order approximation gives

\begin{equation}
e_{I} \approx \frac{\partial I}{\partial R}e_{R} + \frac{\partial I}{\partial N}e_{N}.
\label{eq:17}
\end{equation}

The derivatives are evaluated at the reference reflectances and generally vary among observations. Squaring and averaging this expression introduces contributions from each band and a cross-product involving both errors. Consequently, smaller errors in the individual bands do not guarantee a smaller index error: their signs, co-variation and the index sensitivities also matter. For NDVI and EVI2 over typical vegetation reflectances, the red and NIR derivatives have opposite signs. Same-direction band differences can therefore partly cancel in the index, and a correction that changes their balance can alter that cancellation. This provides a mathematical explanation for possible divergence between band and index outcomes, without establishing that error cancellation caused the observed pattern. Quantifying its contribution would require analysis of the paired band residuals before and after adjustment.

\textbf{Normalized ratios and reflectance magnitude.} NDVI is unchanged when both bands are multiplied by the same positive factor. It remains sensitive to differential red and NIR adjustments, but cannot by itself constrain their absolute magnitudes. NIRv retains NIR reflectance magnitude as well as NDVI \citep{Badgley2017}, while EVI2 contains an additive denominator term and is likewise sensitive to reflectance magnitude \citep{Jiang2008}. For illustration, the operational factors at source NDVI 0.8 change NDVI to approximately 0.815 and increase NIRv by approximately 7.0\%. These are algebraic consequences of the correction, not estimates of improvement relative to OLCI. They explain why a small numerical NDVI response can accompany a larger response in an index that retains reflectance magnitude.

The modest typical change in NDVI agreement therefore neither establishes spectral equivalence nor removes the rationale for adjusting reflectance. Conversely, a larger response in EVI2 or NIRv is not inherently beneficial. The relevant question is whether the transformed values agree more closely with the intended reference under comparable conditions. \citet{Swinnen2022} similarly found that benefits of SRF correction demonstrated with controlled spectral data were not confirmed in their real-sensor evaluation. Their study supports assessing the complete application chain; it does not establish a universal lack of benefit from spectral correction.

\textbf{PPI and its historical reference.} PPI introduces an additional dependence on the historical distribution of the NIR-minus-red difference, as well as illumination geometry \citep{Jin2014}. In the implementation assessed here, the maximum-difference reference is derived from a bounded historical 95th percentile. Adjusting reflectance changes both the current difference and the series used to estimate that reference. Because the transformation is nonlinear, these changes need not reinforce one another or preserve the ranking obtained for an index without a historical reference. In the logarithmic part of PPI, sensitivity to the current difference increases as it approaches the maximum-difference reference, with other parameters fixed. Thus, reference construction and the distance to that reference can materially affect agreement.

The separate historical parameterization of each stream is consequently part of the evaluated PPI calculation. It preserves consistency between each stream's reflectance domain and its reference, but the MODIS and OLCI histories differ in duration and sampled conditions. The observed PPI deterioration cannot therefore be attributed solely to the instantaneous spectral adjustment, nor solely to the reference period. A targeted sensitivity analysis could separate current-band and reference effects by varying them systematically on common observations, with matched-period references expressed in the appropriate reflectance domain. Such an analysis would address attribution; it is not supplied by the present before-and-after statistics.

For vegetation-monitoring applications, the appropriate evaluation follows the variable ultimately used. Reflectance agreement establishes the behavior of the harmonized inputs, index agreement establishes the behavior of their transformation, and phenology or drought-product assessment establishes the consequences for the intended use. Better agreement at one level does not determine accuracy at the next. The present evidence therefore supports an explicit reflectance adjustment with variable-specific assessment, while leaving its effects on phenological timing, seasonal integrals and drought classifications to direct evaluation of those products.

\subsection{Applicability, uncertainty and implications for long-term records}\label{sec:applicability-uncertainty-and-implications-for-long-term-records}

The practical value of direct empirical estimation is a fixed mapping that can be applied using MODIS reflectance alone, including years without OLCI observations. Its scope is defined by the input products and normalization framework. The six-year overlap samples seasonal and interannual variation, and the European assessment supports improved reflectance agreement at most sampled locations. Application elsewhere requires sufficiently comparable products, spatial support and vegetation conditions. Changes to the atmospheric correction, BRDF implementation or reference bands warrant reassessment of the numerical coefficients.

\textbf{Spatial and sampling support.} Calibration on homogeneous land-cover neighborhoods reduces mixing at class boundaries but gives less direct support for fragmented landscapes. Applying the same functions to mixed pixels assumes that their spectral relationships are sufficiently represented by source NDVI, an assumption whose limitations are illustrated by the background dependence in Section 4.1. Reprojection aligns sampling grids without equalizing sensor footprints. Similarly, a 3 × 3 patch provides a local summary rather than a finer-resolution observation or nine independent replicates. Because valid-cell membership can change among dates, a patch time series can also reflect changes in the sampled surface mixture. The paired masks used here ensure a fair before-and-after comparison within each variable, but do not remove that source of temporal variation.

The 460-location assessment gives each site equal weight, so locations with many valid dates do not dominate the continental summary. Nevertheless, the locations are not a documented probability sample, and unavailable inputs at 40 requested locations leave a gap in the assessed coverage. Reported improvement rates therefore describe this sample rather than the proportion of European land area that would improve. Performance by land-cover, climate and landscape heterogeneity would provide more specific guidance about where the correction is useful. Such estimates require sufficient observations within each group and explicit treatment of clustered locations.

\textbf{Uncertainty and transfer.} The uncertainty relevant to an application includes uncertainty in the fitted adjustment, residual disagreement between products and uncertainty introduced by spatial and temporal matching. Variation among annual functions addresses only part of this problem. A narrow range of predicted factors can coexist with appreciable residual disagreement at individual dates or locations. Conversely, a substantial residual need not indicate an incorrect factor if the paired products represent different surface mixtures or temporal observation windows. Reporting the adjustment function alongside site-level errors and data support is therefore more informative than treating coefficient stability as a complete uncertainty estimate.

\textbf{Historical application and record continuity.} Applying a fixed relationship to earlier MODIS observations assumes that the source-to-reference mapping remains adequate beyond the overlap period. Earlier years may differ in calibration, acquisition conditions and the distribution of vegetation states. A fixed spectral adjustment is not a correction for arbitrary time-varying sensor drift. It also need not preserve the magnitude of every trend or anomaly, because its factors change with NDVI. Continuity studies have shown that sensor transitions can introduce artifacts into vegetation trend analyses \citep{Tian2015}, while empirical intersensor transformations can support continuity when their processing conditions and scope are respected \citep{Roy2016}.

Agreement during overlap and continuity across a joined record should consequently be assessed separately. The former compares corresponding observations; the latter asks whether the transition introduces a departure relative to the record's temporal behavior. A change near the transition can also reflect environmental variability, so its timing alone does not establish a sensor artifact. For MR-VPP applications, the next level of evidence is therefore the behavior of the joined vegetation signal and the resulting phenological or drought variables under a documented processing and reference configuration. The present equations provide a reproducible harmonization step, while the assessment identifies the extent to which that step improves the inputs used by those applications.

\FloatBarrier

\section{Conclusions}\label{sec:conclusions}

This study establishes an operational MODIS-to-OLCI adjustment from paired satellite surface reflectances and examines its relationship to adjustment derived from common modeled spectra. Both satellite streams undergo atmospheric correction followed by Ross--Thick/Li--Sparse Reciprocal BRDF normalization to nadir viewing and local solar noon. Six overlapping years provide empirical relationships under this common physical framework. The implemented quadratic red and linear NIR functions use unadjusted MODIS NDVI and can be applied without coincident OLCI observations.

The specified irradiance-weighted canopy simulation does not reproduce the empirical relationship, particularly for red reflectance at high NDVI. Background-dependent simulated ratios also show that NDVI alone does not uniquely determine the adjustment. These results distinguish the information represented by a spectral-convolution calculation from that contained in actual normalized satellite pairs. They establish the need to examine a simulation-derived formulation against the intended products, while the poor pooled quadratic approximation cautions against attributing all discrepancies to the physical model or to satellite processing.

Across 460 European locations during 2019--2024, the empirical adjustment reduced red and NIR RMSE at 66.3\% and 71.1\% of locations, with median paired reductions of 5.98\% and 13.67\%. The corresponding median changes were −0.86\% for NDVI, +0.95\% for EVI2, +0.74\% for NIRv and +7.07\% for PPI. The operational mapping therefore improves reflectance consistency at most assessed locations, while benefits for vegetation indices are less consistent. PPI outcomes include both current reflectance and independently constructed historical reference parameters.

The contribution is a reproducible adjustment grounded in actual normalized observations, evidence of a discrepancy with the specified simulation, and a quantified account of the correction's effects on vegetation signals. For users continuing the MODIS record with OLCI, these results support reflectance harmonization under the documented configuration but do not establish improved PPI or phenological continuity. Separating current-reflectance and historical-reference effects is the next step for that application. A common withheld satellite assessment of simulation-derived and empirical corrections would further establish their relative suitability.

\FloatBarrier

\section{Code and data availability}\label{sec:code-and-data-availability}

This study analyzed existing MODIS and Sentinel-3 NBAR time series; no new observational data were collected. Custom scripts were used for data preparation, residual modelling, calendar-matched contrast construction, and S-score calculation. These scripts are not publicly available.

\FloatBarrier

\section{Author contributions}\label{sec:author-contributions}

Hongxiao Jin: Conceptualization, Methodology, Software, Validation, Formal analysis, Visualization, and Writing - review and editing.

\FloatBarrier

\section{Competing interests}\label{sec:competing-interests}

The author declares no competing interests.

\FloatBarrier

\section{Acknowledgements}\label{sec:acknowledgements}

The simulation and demonstration were made on LUNARC high performance computation system within National Academic Infrastructure for Super­computing in Sweden (NAISS). The test data are from European Environment Agency (EEA) under a service contract to Flemish Institute for Technological Research (VITO).

\FloatBarrier

\section{AI usage}\label{sec:ai-usage}

OpenAI GPT-6.1 Sol was used to assist with literature searches, transfer word document to LaTex, and improvements to language and readability. The author reviewed and verified all AI-assisted outputs and takes full responsibility for the accuracy, integrity and originality of the work.

\FloatBarrier
\bibliographystyle{plainnat}
\bibliography{references}
\end{document}